\documentstyle[12pt,epsf]{article}
\newcommand{\bc}{\begin{center}}
\newcommand{\ec}{\end{center}}
\newcommand{\be}{\begin{equation}}
\newcommand{\ee}{\end{equation}}
\newcommand{\ba}{\begin{eqnarray}}
\newcommand{\ea}{\end{eqnarray}}
\newcommand{\bt}{\begin{tabular}}
\newcommand{\et}{\end{tabular}}

\begin{document}
\begin{flushright}
ITEP-PH 01--99\\
~~~KEK--TH99--610\\
~~~~KIAS--P99009~\\
~~~~SNU 99-007\\[3mm]
~~~~June 17 1999\\
\end{flushright}
\vspace{0.5cm}

\begin{center}
{\Large\bf
The non--perturbative corrections to the $\bar B \to X_s\gamma$ \\[5mm]
photon spectrum in a parton--like model}
\\[1cm]
Y.--Y.Keum$^{(a)}$\footnote{E--mail: keum@ccthmail.kek.jp}, 
P.Yu.Kulikov$^{(b)}$, I.M.Narodetskii$^{(b)}$\footnote{E--mail:
naro@heron.itep.ru}, 
H.S.Song$^{(c)}$\footnote{E--mail: hssong@physs.snu.ac.kr}\\[5mm]
$^{(a)}$ Theory Group, KEK, Tsukuba, Ibaraki, 305-0801 Japan, \\
$^{(b)}$ Institute of Theoretical and Experimental Physics, 117218
Moscow, Russia\\
$^{(c)}$ CTP, Seoul National University, Seoul, Korea\\[2mm]
\end{center}
\vspace{0.5cm}
\begin{abstract}
We derive a new parton--like formula, which establishes a
simple connection between the electroweak decay rate $\Gamma (\bar B \to 
X_s\gamma$) and the rate of a free $b$--quark decay. The main features of our
approach are the treatment of the $b$--quark as an on--mass--shell
particle and the inclusion of the effects arising from the $b$--quark
transverse motion in the $\bar B$--meson. Using various $b$--quark
light--front (LF) distribution
functions, both phenomenological one and the ones derived from current
constituent quark models, and neglecting perturbative corrections 
we compute the photon energy spectra and the moments of the shape 
function. It is shown that the parton--like approach is 
fully consistent with the Heavy Quark Effective Theory (HQET) 
provided the $b$--quark constituent mass is redefined in the way similar 
to that used in HQET to define  the pole mass of the $b$ quark. In this
way the correction to first order in $1/m_b$ can be eliminated from the total 
width in agreement with the general statement of HQET. 
We have also found that the photon energy spectra calculated
in  the LF approach agree well with the ones obtained in the
ACM model, provided the same distribution function is used as input in
both cases. In spite of the simplicity of the model our results show a
fair good agreement both with the HQET predictions and available
experimental
data. 
\end{abstract}

\vspace{0.5cm}


\vspace{0.5cm}


\newpage
\noindent 
Studying of the photon spectrum in the weak radiative $\bar B\to X_s\gamma$ 
is important for understanding how precisely the total rate can be predicted 
in the presence of an experimental cut on the photon energy E. 
Experimentally, because of backgrounds, only 
the high energy part of the photon spectrum 
can be detected with the present experimental cut  
$E> 2.1~{\rm GeV}$ at CLEO \cite{G98}.
Gross features of the spectrum such as the average photon energy 
may be used to measure the fundamental HQET parameters 
\cite{FLS94}, \cite
{KL95}. Since the $b$--quark is heavy compared to the QCD scale, the inclusive 
$B\to X_s\gamma$ decay rate can be calculated in a systematic QCD--based 
expansion \cite{CGG90}. However, near the end point an important
non--perurbative 
effect due to the soft interactions of the $b$--quark with the light constituents 
has to be included. This so--called "Fermi motion" can be included in the heavy 
quark expansion by resumming an infinite set of leading--twist
contributions 
into a shape function $F(x)$ \cite{ShF}, where a scaling variable is defined as 
$x=(2E-m_b)/\bar\Lambda$ with 
$\bar\Lambda=M_B-m_b$.  
This is quite analogous to what happens for the structure function in 
deep--inelastic scattering in the region where the Bjorken variable $x_B\to 1$.
A model independent determination of the shape function is not available
at the present time, 
however it may be possible to address this issue using lattice QCD \cite{A98}.
Ans\"atz for the shape function constrained by the information on its few
first 
moments has been recently used in Ref. \cite{KN99} including the full NLO 
perturbative QCD corrections. 

As to phenomenological analyses of the photon spectra up to now they 
have been solely based on the ACM model \cite{AP79}, \cite{ACM82}. 
This model treats the heavy hadron as a bound state 
of the heavy quark and a spectator, with a certain momentum distribution. 
A light--front (LF) approach to consideration of 
the inclusive semileptonic 
transitions 
was suggested in Refs. \cite{JP94}--\cite{GNST97} and has been recently
refined in 
\cite{KNST97}. 
The corresponding ans\"atz of Ref. \cite{KNST97} reduces to a specific 
choice of 
the primordial LF distribution function $|\psi(\xi,p^2_{\bot})|^2$, 
which represents the probability to find the 
$b$ quark carrying a LF 
fraction $\xi$
and a transfer
momentum squared $p^2_{\bot}=|{\bf p}_{\bot}|^2$. 
As a result, a new parton--like formula for the inclusive semileptonic 
$b\to c,u$ width has been derived \cite{KNST97}, which is similar 
to the one obtained by Bjorken {\it et al.} \cite{BDT92} in case of infinitely heavy  
$b$ and $c$ quarks.

Some of the dinamical features of this model get obscured by the integration 
over the lepton energy. They are seen in a cleaner way in the spectrum of the 
photons in the radiative $\bar B\to X_s\gamma$ transitions. 
In this paper, we extend the work of Ref. \cite{KNST97} to compute 
the non--perturbative corrections to the  photon spectrum and the 
$\bar B\to X_s\gamma$ inclusive rate\footnote{A preview of this work can
be
found 
in \cite{KNST98}.}. We strive to 
implement the binding and the $B$--meson wave function effects on the photon 
energy and the invariant mass distributions of the hadrons recoiling against 
the photon. 
We will also study the comparison between the photon spectra $d\Gamma/dE$ 
calculated in the LF and ACM approaches  
and will show that the discrepancy between the two is very  
small numerically. 

To begin with, we briefly discuss a derivation of the inclusive photon energy spectrum 
for the decays $B\to X_s\gamma$ in the context of the LF  
approach of Ref. \cite{KNST97}. Similar to the ACM model the LF quark 
model treats the beauty meson $B$ as consisting of the heavy $b$ quark plus a 
spectator quark. Both quarks have fixed masses, $m_b$ and $m_{sp}$, though. 
This is at variance with the ACM model, that has been introduced in order to avoid the notion of the 
heavy quark mass at all.
 
The $b\to s\gamma$ transition at the fundamental level is generated 
by electroweak penguins \cite{PNGN}, \cite{CD82}. In the leading
logarithmic approximation 
the decay $b\to s\gamma$ is described by the effective Lagrangian 
\be
\label{1}
{\cal L}_{eff}=\frac{4G_F}{\sqrt{2}}
V_{tb}V_{ts}^*c_7(\mu){\bar m}_b(\mu)O_7,
\ee
where
\be
\label{2}
O_7=\frac{e}{32\pi^2}{\bar u}_s\sigma_{\mu\nu}(1+\gamma_5)u_bF^{\mu\nu}.
\ee
In Eqs. (\ref{1}),(\ref{2}) 
$G_F$ is the Fermi constant, $V_{ij}$ are the elements of the
CKM matrix, 
${\bar m}_b(\mu)$ and $c_7(\mu)$ are running $b$--quark mass and the
Wilson coefficient, respectively, evaluated at a subtraction point $\mu$,
and $F^{\mu\nu}$ is the electromagnetic field strength tensor. The
strange quark mass will be neglected throughout this paper; it only enters 
the final results quadratically as $m_s^2/m^2_b$. For simplicity we neglect 
contributions of other operators in the effective Lagrangian which appear 
at next--to--leading order. 

To calculate the decay rate $\bar B\to X_s\gamma$ we use the approach of Ref. 
\cite{KNST97} which is based on the hypothesis of 
quark--hadron duality. 
This hypothesis assumes that the sum over all possible strange final
states $X_s$ can be modeled by the decay width of an on--shell $b$ quark
into on--shell $c$--quark weighted with the $b$--quark distribution
function $f(\xi,p^2_{\bot})$. 
Going through the intermediate steps (for the detail see Ref.
\cite{
KNST97}) we obtain for the partial decay rate   of the inclusive decay 
$B\to X_s\gamma$  
\be
\label{3}
d\Gamma=4x_0\Gamma_0\int\frac{d^2p_{\bot}d\xi}{\xi}f(\xi,p^2_{\bot})d\tau,
\ee
where 
\be
\label{4}
\Gamma_0=
\frac{\alpha G^2_F}{32\pi^4}|V_{tb}
V_{ts}|^2 c_7^2(m_b) {\bar m}_b(\mu)^2m_b^3
\ee
is the contribution of the matrix element of $O_7$ to the $b\to s\gamma$ 
decay rate, and $x_0=m_b/M_B$. In Eq. (\ref{4}) 
the factor $m_b^3$
comes from the two--body phase space. In HQET $m_b$ is usually
associated with the $b$ quark pole mass $m_{b,pole}$. In our
phenomenological
consideration we associate $m_b$ with the constituent mass of the
$b$--quark, see below. The
normalization of $f(\xi,p^2_{\bot})$ reads 
\be
\label{5}
\pi\int \limits^1_0d\xi\int\limits_0^{\infty}
dk^2_{\bot}f(\xi,k^2_{\bot})=1.
\ee
The factor $1/\xi$ in Eq. (\ref{3}) comes from the normalization of the 
$\bar B \to b\bar d$ vertex \cite{DGNS97}. The phase space factor
$d\tau$ is given by
\be
\label{6}
d\tau=\delta[(p_b-q)^2]EdE.
\ee
We choose the z--axis parallel to the 3--vector {\bf q}, so that $q_+=2E$, 
$q_-=0$, where $q_{\pm}=q_0\pm q_z$, then $d\tau$ 
takes the form 
\be
\label{7}
d\tau=\delta(m^2_b-\frac{p^2_{\bot}+m^2_b}{p_b^+})EdE.
\ee
In the first approximation we neglect $p^2_{\bot}$ in
the argument of the
$\delta$--function. Then, introducing the scaling variable $y=2E/M_B$, the
photon spectrum $d\Gamma/dE$ in $\bar B\to
X_s\gamma$ takes the simple form
\be
\label{8}
\frac{1}{\Gamma_0}\cdot \frac{d\Gamma(\bar B \to 
X_s\gamma)}{dy}=R_{LF}(y),
\ee
where
\be
\label{9}
R_{LF}(y)=\frac{1}{x_0}y{\tilde f}(y),
\ee 
and $\tilde f(\xi)=\pi\int\limits^{\infty}_0
dp^2_{\bot}f(\xi,p^2_{\bot})$. 
Equivalently, one can write the spectrum in the standard QCD form
\cite{BSUV94} 
\be
\label{10}
\frac{1}{\Gamma_0}\frac{d\Gamma}{dE}=\frac{2}{\bar \Lambda}F(x).
\ee
Therefore the specific choice adopted for $\tilde f(\xi)$ corresponds to a 
particular form of the QCD shape function $F(x)=(\Lambda/M_Bx_0)y\tilde
f(y)$, where
$x=(y-x_0)/(1-x_0)$.\footnote{If one uses the Infinite Momentum Frame
prescription \cite{JP94}, \cite{MTM96} $p_b=\xi P_B$, i.e.
$m_b^2(\xi)=\xi^2M_B^2$, then from Eqs.
(\ref{4}),(\ref{6}) one easily derives $d\Gamma\propto E^3{\tilde
f}(y)dE$ \cite{J99}.}

Following Ref. \cite{KNST97} we now account for the transverse motion
of the $b$--quark 
in Eq. (\ref{7}) to find a bit more complicated expression for $R_{LF}(y)$ 
\cite{KNST98}
\be
\label{11}
R_{LF}(y)=2m_b\pi\int^1_y f(\xi,p^{*2}_{\bot}) d\xi,
\ee
where the integration limits follow from the condition $p^{*2}_{\bot} \ge 
0$,
with $p^{*2}_{\bot}=p^{*2}_{\bot}(\xi,E)=m^2_b(\xi/y-1)$. In this case the
shape of the spectrum is obtained by direct integration
of the distribution 
function. 
We shall see below that the difference between $R_{LF}(y)$ given by Eqs. 
(\ref{9}) and (\ref{11}) is very small numerically.

In the free quark approximation,  
$f(\xi,p^2_{\bot})=\delta(\xi-\xi_0)\delta(p^2_{\bot})$, the total
inclusive 
width $\Gamma(\bar B\to X_s\gamma)$ 
is the same as the radiative $b\to s\gamma$ width $\Gamma_0$, and the spectrum of photons 
is a monochromatic line:
\be
\label{12}
\frac{d\Gamma_0}{dE}=\Gamma_0\delta(E-\frac{m_b}{2}).
\ee
The delta--function of Eq. (\ref{12}) is transformed into a peak of a 
finite width due to the heavy quark motion. This effect is solely responsible
 for the filling in the windows between $m_b/2$ and the kinematical boundary 
in the $B$ meson decay, $E_{max}=M_B/2$.
\footnote{The true endpoint is actually located at 
$[M_B^2-(m_K+m_{\pi})^2]/2M_B\approx 2.60~{\rm GeV}$, i.e. slightly below 
$M_B/2\approx ~2.64~{\rm GeV}$.}. The expressions 
in Eqs. 
(\ref{9}), (\ref{11}) exhibit a pronounced peak which is rather assymetric. 
It is gratifying feature of the LF model, since it is in qualitative accord 
both with findings in QCD and experimental data.
The perturbative corrections 
arising from gluon Bremsstrahllung and one--loop effects \cite{AC} also lead 
to a nontrivial photon spectrum at the partonic level. Since our primary 
object here is to discuss non--perturbative effects due to the Fermi motion, 
we will implicitly ignore perturbative gluon emission throughout our analysis.
In this case the parton matrix element squared is a constant and can be 
taken out of the integral in Eq. (\ref{3}).

Since we do not have an explicit representation for the B--meson Fock 
expansion in QCD, we shall proceed by making an ans\"atz for the momentum 
space structure of the wave function. 
This is model dependent enterprise but has its close equivalent in 
studies of $\bar B \to X_s\gamma$ using the ACM model.
In what follows, we
will adopt both a phenomenological LF wave function and the LF
functions corresponding to the various equal time (ET)
quark model wave functions. As to the phenomenological ans\"atz, 
we use a model  first written in Ref. \cite{DK93}, and
also 
employed in Refs. \cite{MTM96},\cite{GNST97} to implement the bound state
effects in 
$B$--meson decays. It is written 
in the Lorentz--invariant form
\be
\label{13}
\psi(|{\bf p}|)={\cal N}exp(-\frac{\lambda }{2}v_Bv_{sp})={\cal N}exp (-\frac{\lambda}{2} \frac
{\varepsilon_p}{m_{sp}}),
\ee
where $v_B$ and $v_{sp}$ are the 4--velocities of the $b$--quark and the 
quark--spectator, 
respectively, 
and $\varepsilon_p=\sqrt{|{\bf p}|^2+m_{sp}^2}$ is the 
energy of the spectator. We shall use the normalization condition
$\int \limits ^{\infty}_0 p^2dp\psi^2(|{\bf p}|)/2\varepsilon_p=1$,
in which case ${\cal N}^2=2\lambda/(m_{sp}^2K_1(\lambda))$,
where $K_1(\lambda)$ is the McDonald function. The function $\Phi(p^2)=
\psi^2(|{\bf p}|)/2\varepsilon_p$ represents a momentum distribution of 
the spectator quark in the $B$ meson rest frame.
We convert from ET to LF momenta by leaving the transverse momenta
unchanged and letting 
\be
\label{14}
p_{iz}=\frac{1}{2}(p_i^+-p_i^-)=\frac{1}{2}(p_i^+-\frac{p^2_{i{\bot}}
+m^2_i}{p_i^+})
\ee 
for both the $b$--quark $(i=b)$ and the quark--spectator
$(i=sp)$.
The longitudinal LF momentum fractions $\xi_i$ are defined as $\xi_{sp}=
p^+_{sp}/P_B^+$, $\xi_b=p^+_b/P_B^+$, 
with $p^+_b+p^+_{sp}=P^+_B$. In the $B$--meson rest frame $P_B^+=M_B$.
Then for the distribution function 
$f(\xi, p^2_{\bot})$ ($\xi=\xi_b$) normalized according to
Eq.(\ref{5}) one obtains
\be
\label{15}
f(\xi,p^2_{\bot})=\frac{1}{8\pi}\cdot \frac
{\psi^2(\xi,p^2_{\bot})}{1-\xi}=
\frac{{\cal N}^2}{8\pi\cdot (1-\xi)}\exp\left[-\frac{\lambda}{2}\left(
\frac{1-\xi}{\xi_0}+\frac{\xi_0}{1-\xi}
(1+\frac{p^2_{\bot}}{m_{sp}^2}\right)\right],
\ee
where $\xi_0=m_{sp}/M_B$. The function (\ref{15}) is sharply peaked at 
$p^2_{\bot}=0$, $\xi=\xi_0$. In what follows, we shall refer to the LF
wave function of Eq. (\ref{15}) as the case A.

A priory, there is no connection between the ET momentum distribution 
$\Phi(p^2)$ of a constituent quark model and LF wave function
$\psi(x,p^2_{\bot})$. However, the mapping between 
the variables described above turns a normalized solution 
of the ET equation of motion into a normalized solution of the different
looking LF equation \cite{Co92}. Because the ET function depends on the
relative momentum 
it is more convenient to use the quark--antiquark rest frame instead of the 
$B$--meson rest frame. Recall that in the LF formalism these two 
frames are different. As a result one obtains 
the LF wave function as $\psi(\xi,p^2_{\bot})=(\partial p_z/\partial
\xi)\Phi(p^2_{\bot}+p^2_z(\xi,p^2_{\bot}))$. Explicit form of this
function
is given {\it e.g.} 
by Eq. (10) of Ref. \cite{GNST97}. It is wave functions made kinematically 
relativistic in this fashion, that were used in a recent calculation of
the $B_c$ lifetime \cite{ANSS99}. We calculate the photon energy spectra 
using the three  representative LF wave functions corresponding to the
non--relativistic ISGW2 \cite{SI95}, AL1 \cite{SS94}, and relativized 
DSR \cite{SS97} constituent quark models\footnote{The ET ISGW2 function 
corresponds to the Gaussian distribution 
$\Phi (p^2)$ conventially employed in the ACM model with $p_F=0.43~{\rm GeV}$.
For AL1 and DSR models we use simple analytical
parametrizations of the ET wave functions 
\cite{SS99}.}. The main difference between the ET wave functions of these
models relies in the behaviour at high value of the internal momentum, for
further discussion see \cite{ANSS99}. We believe that the spread of
results obtained for these 
distribution functions is a fair representation of model dependence 
resulting from the inclusion of Fermi motion.

Having specified the non--perturbative aspects of our calculations, we 
proceed to present numerical results for the photon spectrum in the decay 
$B\to X_s\gamma$. In case A we take 
$m_b=4.8~ {\rm GeV}$, $m_{sp}=0.3~ {\rm GeV}$ and 
$\lambda=2$ as reference values. These values are motivated by a study 
of the $b\to c$ decays \cite{GNST97}. 
For the models B to D we use the constituent quark masses listed in Table 1.

The choice of $m_b$ in our approach deserves some comments. The
numerical calculations using the constituent $b$--quark masses
show large deviations of the $\Gamma(\bar B \to X_s\gamma)$ from the free 
decay rate $\Gamma_0\propto m_b^3$ ($\approx 10\%$ for the cases B to D,
see Table 2). This signals
the appearence of the linear $1/m_b$ corrections to the free quark limit.
The reason is that  the 
constituent quark models usually employ the $b$--quark 
masses that are $300-400~{\rm MeV}$ higher than the pole $b$--quark mass. 
This fact seems to be a subtlety in applying 
the constituent quark model approach to calculate the
non--perturbative corrections to the $\bar B\to X_s\gamma$ inclusive rate.
To overcome the uncertainties induced by the  constituent mass of 
the $b$--quark we use a simple phenomenological receipe that
considerably improves the situation. Notice that, as in the
ACM model \cite{BSUV94}, $1/m_b$ corrections can be absorbed into
the
definition of the $b$--quark mass. We introduce 
${\tilde m}_b=m_b+\delta m_b$ by imposing the condition 
$\bar y({\tilde m}_b)=x_0$, where ${\bar y}=\int\limits_0^1yR_{LF}(y)dy$. 
This condition coincides with that used in HQET to define the 
pole mass of the $b$--quark. 
As a result,
the correction to first order in $1/m_b$ will be eliminated from the
total 
width in agreement with the general statement of HQET.
To illustrate our arguments, consider the analytically
tractable case of the photon spectrum of Eq. (\ref{9}) with the 
distribution function given by Eq. (\ref{15}). In this case 
$\tilde f(\xi)=2\lambda/(\xi_0K_1(\lambda))
\exp\left(-\lambda/2[\xi_0/(1-\xi)+(1-\xi)/\xi_0]\right)$, 
and simple analytical  
expressions for $R_{LF}=\int\limits^1_0R_{LF}(y)dy$ and $\bar y$ are
avaiable,
\be
\label{16}
R_{LF}=\frac{1}{x_0}(1-\xi_0\kappa_2),~~~
\bar y=\frac{1}{x_0}(1-2\xi_0\kappa_2+\xi_0^2\kappa_3),
\ee
where $\kappa_n=K_n(\lambda)/K_1(\lambda)$. The $1/M_B$ correction to
$R_{LF}$ 
can be absorbed into the definition of ${\tilde m}_b$. Indeed, 
neglecting $\xi^2_0$ and letting $\tilde m_b\approx
(1-\xi_0\kappa_2)M_B$, 
one obtains 
$R_{LF}=1+O(1/M_B^2)$. 
The $B$--meson mass can be eliminated
in favour of the ¤b¤--quark mass, so we have the desired result, 
$R_{LF}=1+O(1/m_b^2)$. 

We have calculated numerically the values of ${\tilde m}_b$ in
different models using Eq. (\ref{11}) for $R_{LF}(y)$. Although 
${\tilde m}_b$ depend on the assumed shape of distribution, this
dependence
is marginal: the uncertainty on $\tilde m_b$ is between 4.6 and 4.7
GeV, depending on the choice of $f(\xi,p^2_{\bot})$, (see Table 1). These
values are 
consistent whith the $b$ quark pole mass $m_{b,pole}=4.8\pm 
0.15~{\rm GeV}$ \cite{N94}. If we repeat the same exercise by applying Eq.
(\ref{9}) we find practically identical values of ${\tilde m}_b$.

We first study the photon spectra using Eqs. (\ref{9}) and
(\ref{11}).  
Our results for the photon spectra and moments are reported in Fig. 1 
and Table 2. The different curves in Fig.1 correspond to the 
models A to D. For each case we show separately the spectra calculated
from Eqs. (\ref{9}) and (\ref{11}) using both $m_b$ and $\tilde m_b$. 
The influence 
of the various choices of the distribution function and the $b$--quark
mass can be read off from Table 2, 
where we show both $R_{LF}(m_b)$, $R_{LF}({\tilde m}_b)$ and analogous 
quantities $R_{ACM}$ calculated in the ACM model, see below. 
In Table 2 we also show the average photon 
energy $\bar y$ normalized to $M_B/2$, and the moments ${\bar {y^2}}$ 
and ${\bar {y^3}}$. Here, and henceforth, we present the results
obtained using Eq. (\ref{11}). The substitution $m_b\to \tilde m_b$
modifies the predictions for the total rate and the moments $\bar y$,
$\bar {y^2}$ by
about $8-9\%$ for the models B--D, while it affects the corresponding
quantities by less than $1.5\%$ for the model A. 
Our final results for the integrated photon spectra $R_{LF}(\bar m_b)$
agree well with  the corresponding OPE prediction \cite{FLS94} 
$R_{{\rm OPE}}=1+(\lambda_1-9\lambda_2)/2m^2_b$,
where 
$\lambda_1$ and $\lambda_2$ parameterize the matrix elements of the kinetic 
and chromomagnetic operators, respectively. 
For $\lambda_1=-0.3\pm 0.2~{\rm GeV^2}$ and $\lambda_2=0.12\pm 0.02~
{\rm GeV^2}$ $R_{{\rm OPE}}$ affects the free quark result by a few
per cent, $ R_{{\rm OPE}}=0.975\pm 0.005$.
 
The dependence of the energy moments ${\bar{y^n}}$ on $m_b$ is rather
weak, in contrast to that of the QCD moments of the shape function
$<x^n>=\int \limits^1_{-2M_B/\bar{\Lambda}}x^nF(x)dx,$
which are very sensitive to the difference between $M_B$ and $m_b$.
In particular, changing 
the $b$--quark mass from $m_b$ to $\tilde m_b$ modifies 
$<x^2>$ and $<x^3>$ dramatically. Note that the resulting values of
$<x^2>$ and $<x^3>$ still have a sizable model dependence. Our predictions
for $<x^2>$ are somewhat small, although in agreement (for the model D) 
with 
the result of Ref. \cite{KN99} and compatible with the results obtained
from the QCD sum rules, $<x^2>\approx 0.5$. This means that the LF ansatz
can be made consistent with the QCD description provided the spectator
quark
is relativistic. This conclusion agrees with that of Ref. \cite{BSUV94}. 

In order to compare our results with those of the ACM model we have 
calculated the inclusive $\bar B\to  X_s\gamma$ photon spectra
in a simplified ACM model \cite{BSUV94}  assuming the monochromatic
distribution (\ref{12}) 
for the free $b$--quark. We have used the momentum distributions 
$\Phi(p^2)$ of the spectator quark for the models A to D. 
In all cases we have found that the spectra calculated in ACM and LF 
parton models are almost identical. This is not surprising because we have checked numerically 
that the quark masses $\tilde m_b$ defined using the LF models practically 
coincide with the values of the floating $b$ quark mass 
$m_b^f=M_B-\sqrt{m_{sp}^2+p^2}$ averaged over the distribution $\Phi(p^2)$.  
The integrated energy spectra $R_{ACM}$ for the models A to D are reported
in Table 2, they coincide with $R_{LF}$ within a per cent accuracy.

Although we do not consider here perturbative corrections, it is instructive 
to compare theoretical predictions for the Doppler 
shifted spectrum $dB/dE_{lab}$ in the laboratory frame with the CLEO data.
To perform a fit to the data, we rebin the boosted photon spectra in the
same
energy intervals as used by CLEO and, for each choice of the distribution
function, adjust the overall normalization to give the best fit to the data. 
The results are reported in Table 3, and the best fits are displayed in
Fig.2.  All fits have $\chi^2/n_{dof}\ll 1$, indicating the present
accuracy of experiment. Averaging over the models we obtain
\be
\label{17}
B(\bar B\to X_s\gamma)=(2.5\pm 0.5_{exp}\pm 0.3_{model})\times 10^{-4},
\ee 
where the last error comes from the model dependence. This result is 
consistent both 
with the update CLEO measurement \cite{CLEO98}
$B(\bar B\to X_s\gamma)=(3.15\pm 0.35\pm 0.32\pm 0.26)\times 10^{-4}$, 
and a recent reanalysis \cite{KN99} of the CLEO data
$B(\bar B\to X_s\gamma)=(2.62\pm 0.60_{exp}$$^{+0.37}_{-0.30~th})\times
 10^{-4}$.

Finally, we note that the invariant mass $M_H$ of the hadronic final state 
is related with the scaling variable $y$ by $M^2_H=M^2_B(1-y)$.  Therefore
the 
theoretical results for the photon spectrum can be translated into predictions for 
the hadronic mass spectrum. In Fig. 3 we show the invariant mass
distribution of the hadrons recoiling against the photon for 
the models A--D. Our predictions for hadronic mass spectra must be understood
in the sence of quark--hadron duality. The true hadronic mass spectrum for
low $M_H^2\propto M^2_{K^*}$ may have resonance structure that looks
rather different from our predictions. A realistic model for the hadronic
mass spectrum consists of a single peak located at the mass $K^*(892)$
followed by a continuum (above a threshold value $M_{th}$) which is given
by the inclusive spectrum and is dual to a large number of overlapping
resonances. 
In Table 3 we show the ratios $R_{K^*(892)}=B(\bar B\to K^*\gamma)/
B(\bar B\to X_s\gamma)$ 
obtained by the
integration of the inclusive spectrum in the range $M_H\le
M_{th}$. The result crucially depends on the choice of $M_{th}$; we use 
the value of $M_{th}=1.15~{\rm GeV}$  adopted in Ref. \cite{KN99}. Averaging
over the different model predictions we obtain
$R_{K^*(892)}=0.157^{+0.24}_{-0.44}$. 
This result agrees both with other theoretical predictions and with 
the CLEO measurement  $R^{{\rm exp}}_{K^*(892)}=0.17\pm 0.08$. 
 
In conclusion, 
we have derived a new parton formula, which establishes a
simple connection between the electroweak decay rate $\Gamma (\bar B \to 
X_s\gamma$) and the rate of a free $b$--quark decay. The main features of our
approach are the treatment of the $b$--quark as an on--mass--shell
particle and the inclusion of the effects arising from the $b$--quark
transverse motion in the $\bar B$--meson. Our main result is Eqs.
(\ref{9}), (\ref{11}). Using various $b$--quark distribution
functions we have calculated the photon energy spectra and the corrections 
to the free decay rate. We have shown that the decay width has no linear
to $1/m_b$ corrections only if expressed not in terms of the constituent
quark mass but in terms of a mass $\tilde m_b$ which is defined 
in the way similar to that used in HQET to define  $m_{b,pole}$.
In this way one avoide an otherwise large (and model dependent)
correction of order $1/m_b$ but at expence of introducing the shift in
the constituent quark mass which largely compensates the bound state
effects. A summary of our results presented in Table 2 shows a 
fair good agreement both with the QCD results and avaiable experimental
data. We have also found that the photon energy spectra calculated
in  our LF parton--like approach agree well with the ones obtained in the
ACM model, provided the same ET distribution function $\Phi(p^2)$ is
used as input in both cases. Finally, we note that it would be
interesting to check whether the effective values of the $b$--quark mass
${\tilde m_b}$ can appear to be approximately the same for
different channels ($b\to c$ vs. $b\to u$ or $b\to s$) and for different
beauty hadrons. This work is in progress, and the results will be reported
elsewhere. \\[3mm]

We thank S.Ya. Kotkovsky for the collaboration in the early stages of 
this work. Y.Y.K. thanks M.Kobayashi for his encouragement. Y.--Y.K. and
I.M.N. thank CTP and KIAS for hospitality while part of this work 
was completed. This work was supported in part by the INTAS-RFBR grant No. 
95--1300, INTAS grant No 96--155 and RFBR grant No 95--02--0408a. The work 
by Y.--Y.K. was supported in part by BSRI Program, Project 
No BSRI--97--2414 and the Grant--in--Aid of Japanese Ministry of Education, Science, 
and Sport.

\vspace{1cm}


\newpage
\noindent {\bf Table 1 }. The values of the constituent quark mass $m_b$ and 
$m_{sp}$ (in units of GeV) for the models A to D. The values of $\tilde  m_b$ 
as defined by the 
HQET condition $<x>=0$ are also indicated. 
\vspace{0.5cm}

{\Large\begin{center}
\begin{tabular}{|c|c|c|c|c|}
\hline\hline
Model & A & B & C & D \\
\hline
$ m_b $ & 4.80 & 5.20 & 5.227 & 5.074 \\
\hline
$ \tilde m_b $ & 4.73 & 4.68 & 4.73 & 4.60  \\
\hline
$m_{sp}$ & 0.30 & 0.33 & 0.315 & 0.221  \\
\hline
\hline
\end{tabular}
\end{center}}

\vspace*{0.5cm}

\noindent 
{\bf Table 2. } The total integrated rates $R_{LF}$, $R_{ACM}$,
the moments ${\bar y^n}$, $n=1,2,3$, and the HQET moments  
$<x^2>$, $<x^3>$ of the photon spectra for the cases A to D calculated
using Eq. (\ref{11}). \\[0.5cm] 
{\Large\begin{center}
\begin{tabular}{|c|c|c|c|c|}
\hline\hline
Model & A & B & C & D \\
\hline
$R_{LF}(m_b)$ & 0.973 &0.900 & 0.905 & 0.899 \\
\hline
$R_{LF}(\tilde m_b) $ & 0.987 &0.986 & 0.989 & 0.974 \\
\hline
\hline
$R_{{\rm ACM}}(\tilde m_b)$& 1.008 & 1.005 & 1.008 & 1.003   \\
\hline
\hline
${\bar y}(m_b)$ & 0.873 & 0.806 & 0.818 & 0.795 \\
\hline
${\bar y}(\tilde m_b)$ & 0.885 & 0.873 & 0.885 & 0.851  \\
\hline
${\bar y}^2(m_b)$ & 0.787 & 0.726 & 0.742 & 0.709  \\
\hline
${\bar y}^2(\tilde m_b^2)$ & 0.798 & 0.778 & 0.796 & 0.751  \\
\hline
${\bar y}^3(m_b)$ & 0.712 & 0.655 & 0.676 & 0.636  \\
\hline
${\bar y}^3(\tilde m_b)$ & 0.722  & 0.695 & 0.718 & 0.669   \\
\hline\hline
$<x^2>$ & 0.378 & 0.295 & 0.302 & 0.456  \\
\hline
$<-x^3>$ & 0.348 & 0.130 & 0.176 & 0.348  \\
\hline\hline
\end{tabular}
\end{center}
}

\newpage
\noindent {\bf Table 3. } The branching ratios $B(\bar B \to X_s\gamma)$ obtained
from the fit to 
the CLEO data and the partial
fractions $R_{K^*(892)}$.\\[0.5cm] 
{\Large\begin{center}
\begin{tabular}{|c|c|c|c|c|}
\hline\hline
Model & A & B & C & D \\
\hline
$\Gamma_0/\Gamma_B\cdot 10^4$ & 2.54 & 2.59 & 2.50 & 2.91 \\
\hline
$R_{K^*(892)} $ & 0.1681 & 0.1312 & 0.1486 & 0.1807  \\ 
\hline\hline  
\end{tabular}
\end{center}
}
\vspace{2cm}

\section*{Figure Captions}
{\bf Fig.1}. Theoretical predictions for the photon energy spectrum
using the LF quark models described in the text. Each plot shows the
spectra calculated using both the $b$--quark masses $m_b$ (solid lines)
and $\tilde m_b$ (dashed lines). The thick and thin curves show the
results
obtained using Eq. (\ref{11}) and (\ref{9}), respectively. \\[2mm]
\noindent {\bf Fig.2}. Theoretical predictions for the photon energy
spectrum in the laboratory frame for the different distribution functions
$f(\xi,p^2_{\bot})$ and the corresponding ${\tilde m}_b$. The solid,
dashed, dot--dashed, and dotted curves correspond to the models
A to D, respectively. The data points show the CLEO results. Both the
left--hand plot and the right--hand plot show the results of the best fit
reported in Table 3. \\[2mm] 
\noindent {\bf Fig.3}. Theoretical predictions for the invariant
hadronic mass spectrum for different distribution functions 
$f(\xi,p^2_{\bot})$. 
The notations are the same as in Fig. 2.

\newpage 
\begin{minipage}[t]{10.cm}
\begin{center}
\includegraphics{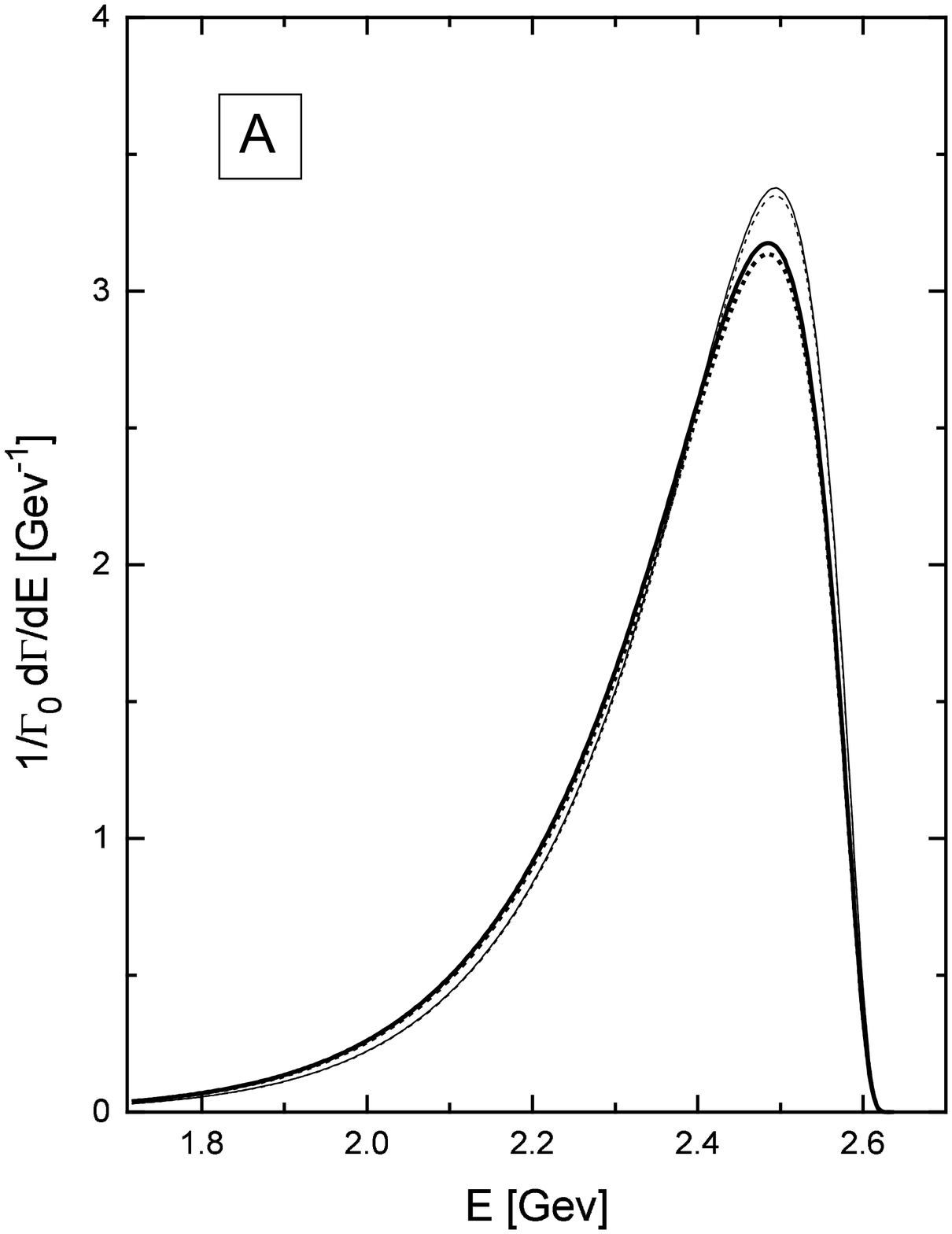}
\includegraphics{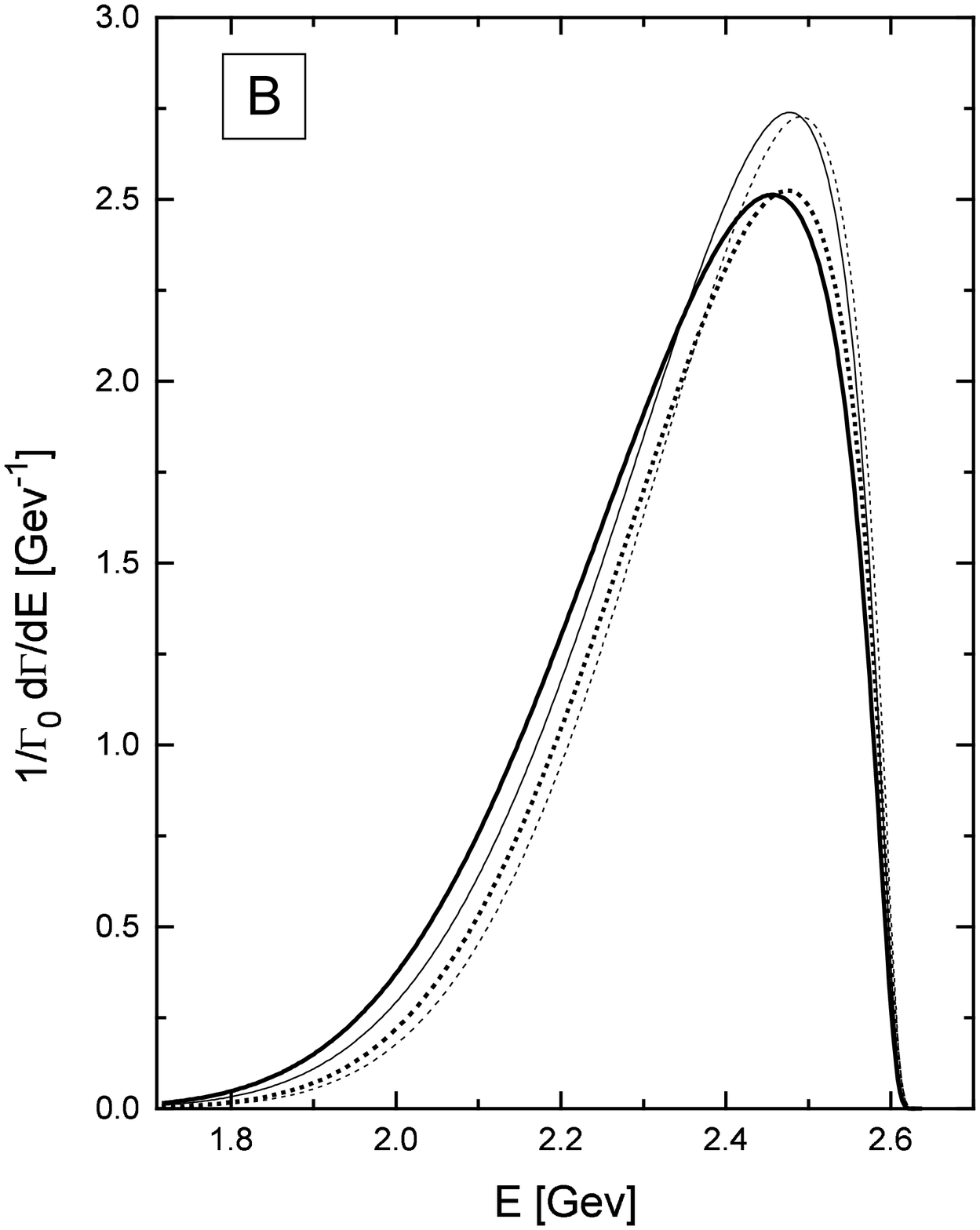}
\includegraphics{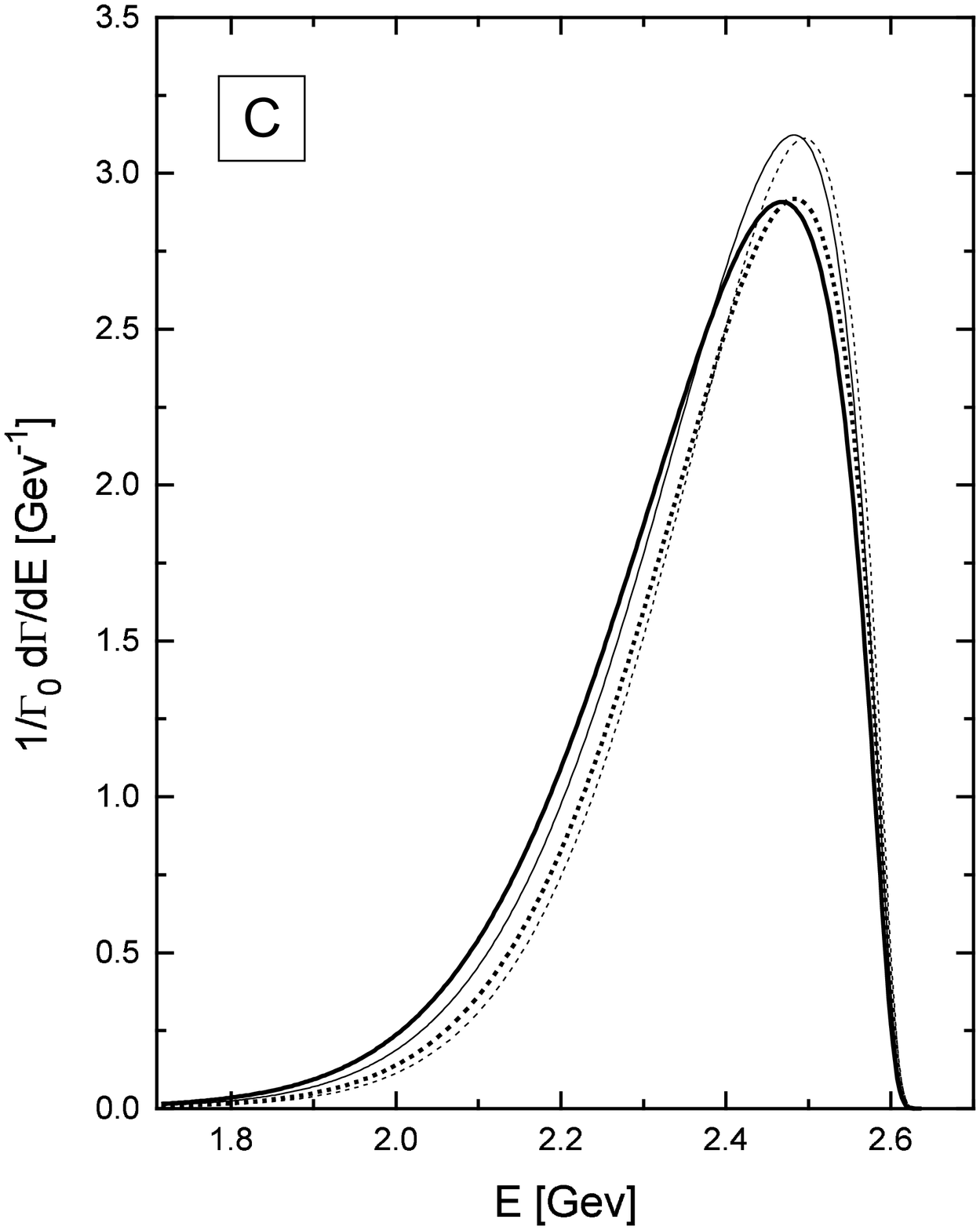}
\includegraphics{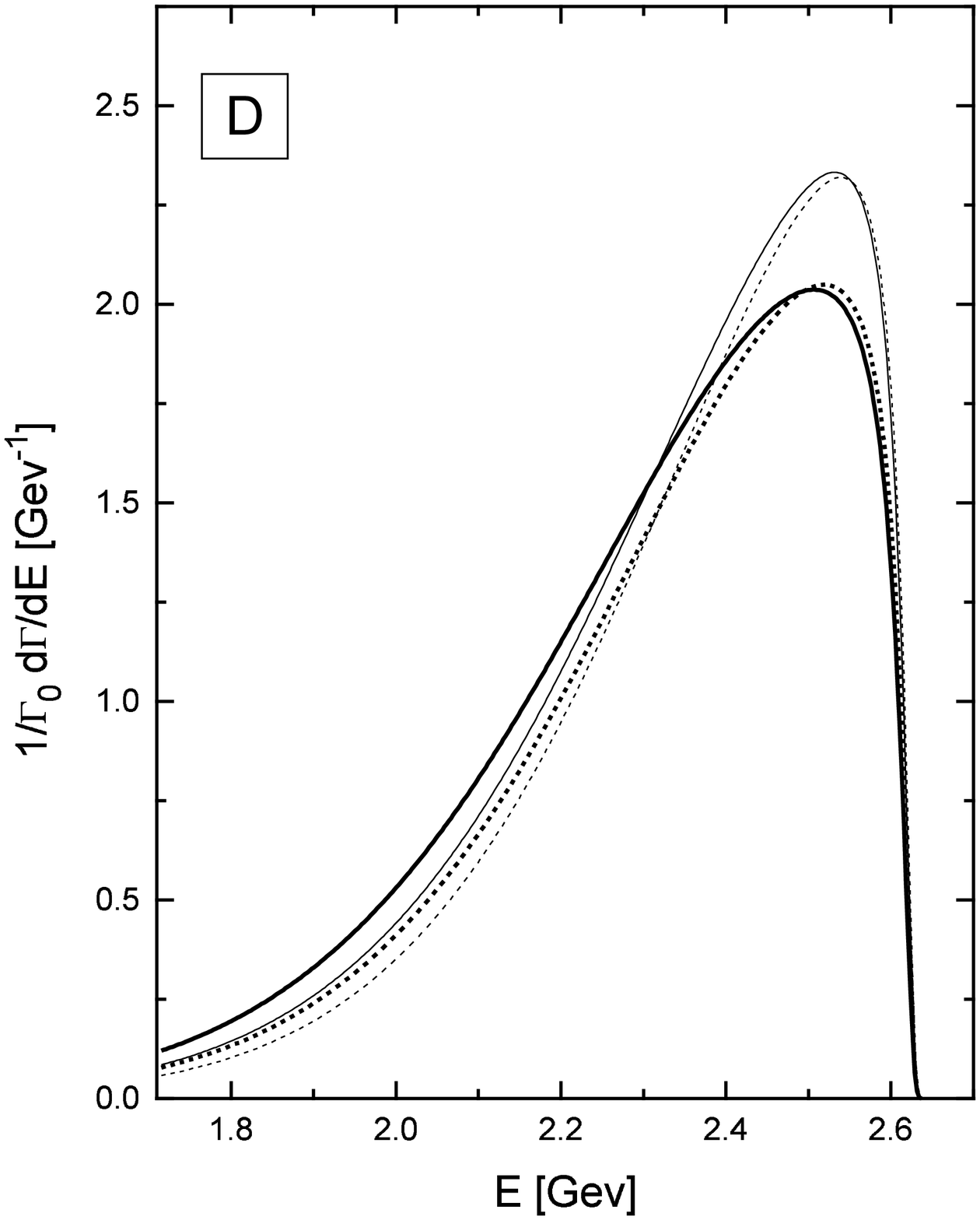}
\end{center}
\vspace*{0.8cm}
\end{minipage}
\vskip 20cm
\bc
{\bf Figure 1} 
\ec
\newpage
\begin{minipage}[t]{10.cm}
\begin{center}
\vskip 0.0cm
\includegraphics{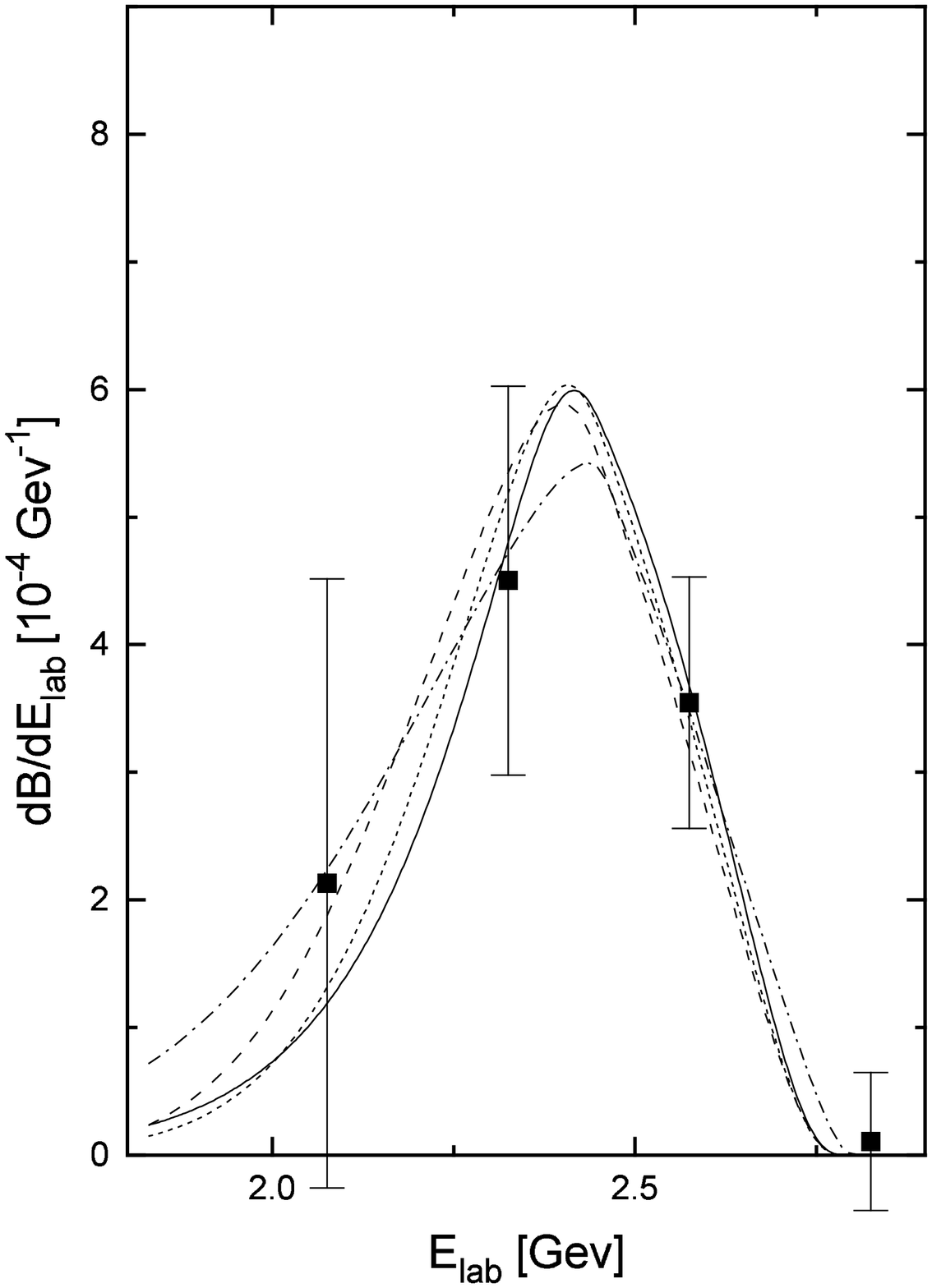}
\includegraphics{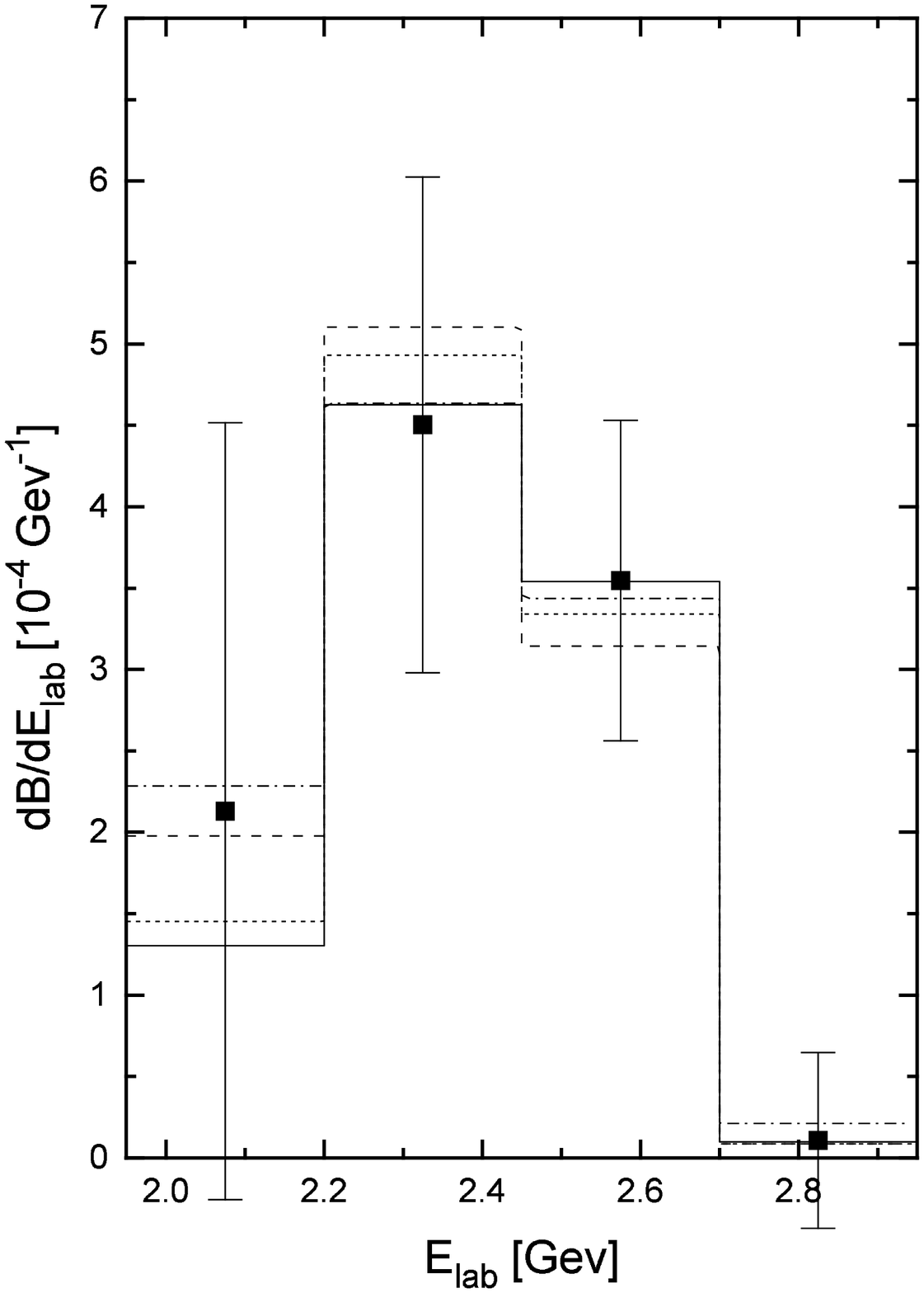}
\includegraphics{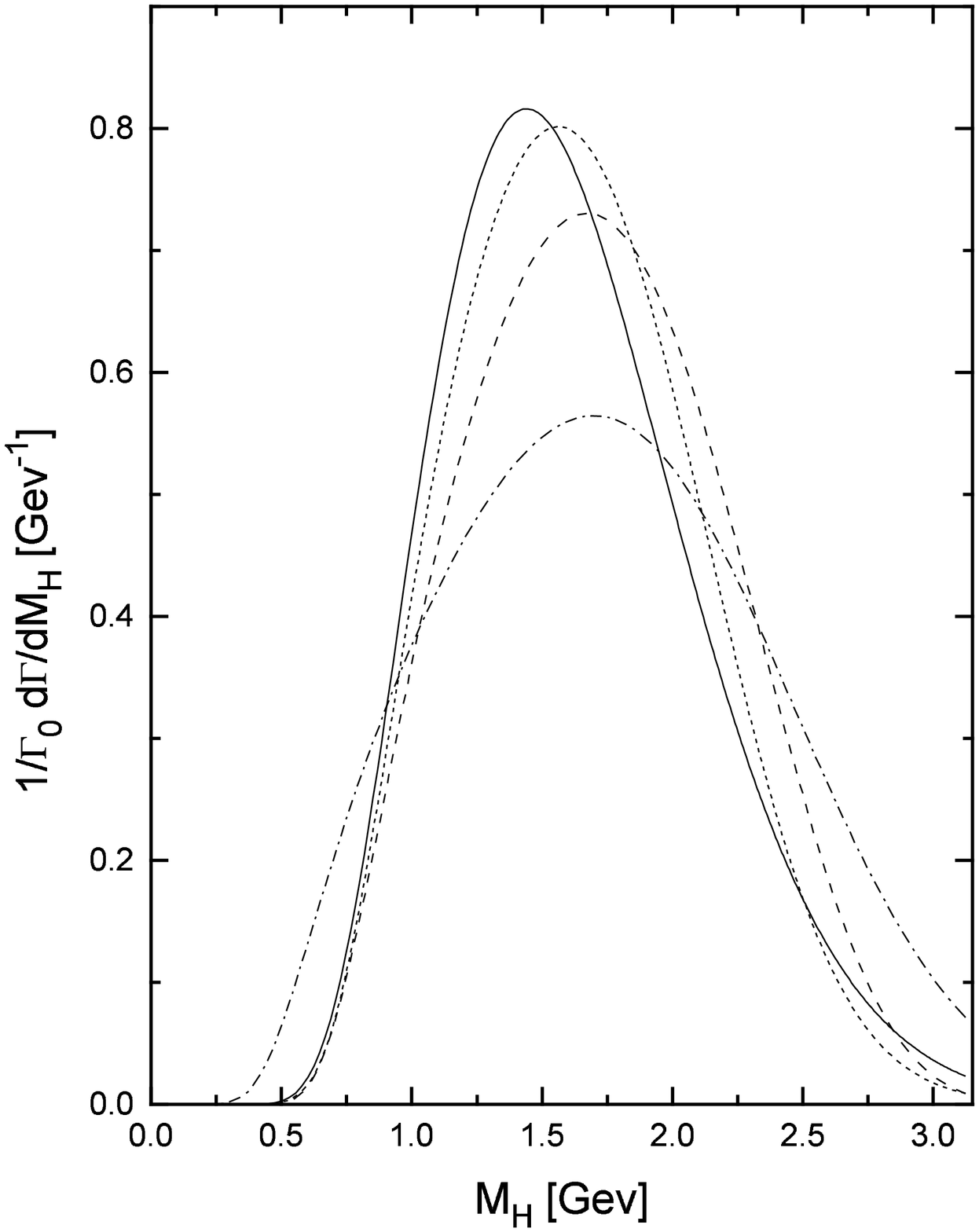}
\end{center}
\vspace{0.3cm}
\end{minipage}
\bc
\vskip 8cm
{\bf Figure 2}.
\vskip 12.5cm
{\bf Figure 3}.
\ec
\end{document}